\documentclass[10pt,letterpaper]{article}
\usepackage[top=0.85in,left=2.75in,footskip=0.75in]{geometry}

\usepackage{changepage}

\usepackage[utf8]{inputenc}

\usepackage{textcomp,marvosym}

\usepackage{fixltx2e}

\usepackage{amsmath,amssymb}

\usepackage{cite}

\usepackage{nameref,hyperref}

\usepackage[right]{lineno}

\usepackage{microtype}
\DisableLigatures[f]{encoding = *, family = * }

\usepackage{rotating}

\usepackage{epsfig,bm,epsf,graphics}
\usepackage{graphicx}

\usepackage{dcolumn}

\raggedright
\usepackage[aboveskip=1pt,labelfont=bf,labelsep=period,justification=raggedright,singlelinecheck=off]{caption}

\makeatletter
\renewcommand{\@biblabel}[1]{\quad#1.}
\makeatother

\date{}

\usepackage{lastpage,fancyhdr,graphicx}
\usepackage{epstopdf}
\usepackage{color}
\fancyheadoffset[L]{2.25in}
\fancyfootoffset[L]{2.25in}

\newcommand{\rev}{\textcolor{black}}
\newcommand{\jj}{\textcolor{black}}

\begin{document}
\vspace*{0.35in}

\begin{flushleft}
{\Large
\textbf\newline{Design Principles of Pancreatic Islets: Glucose-dependent Coordination of Hormone Pulses}
}
\newline
\\
Danh-Tai Hoang\textsuperscript{1,2},
Manami Hara\textsuperscript{3},
Junghyo Jo\textsuperscript{1,4,*}
\\
\bigskip
\bf{1} Asia Pacific Center for Theoretical Physics, Pohang, Gyeongbuk 36763, Korea
\\
\bf{2} Department of Science, Quang Binh University, Dong Hoi, Quang Binh 510000, Vietnam
\\
\bf{3} Department of Medicine, The University of Chicago, Chicago, IL 60637, United States of America
\\
\bf{4} Department of Physics, Pohang University of Science and Technology, Pohang, Gyeongbuk 36763, Korea
\\
\bigskip

%
%





* jojunghyo@apctp.org

\end{flushleft}
\section*{Abstract}
Pancreatic islets are functional units \rev{involved in} glucose homeostasis.
The multicellular system \rev{comprises} three \rev{main} cell types\rev{;}
\jj{$\beta$ and $\alpha$} cells reciprocal\rev{ly} decrease and increase blood glucose by producing \jj{insulin and glucagon} pulses,
while the role of $\delta$ cells is less clear.
Although their spatial organization and \rev{the} paracrine/autocrine interactions between them have been extensively studied, the functional implications of the design principles \rev{are still lacking}.
In this study, we formulated a mathematical model that integrates \rev{the} pulsatility of hormone secretion \rev{and the} interactions and organization of islet cells and examined the effects of different cellular compositions and organizations in mouse and human islets. 
\rev{A common feature of both species was that islet cells produced} synchronous hormone pulses \rev{under low- and high-}glucose conditions,
while they produced asynchronous hormone pulses \rev{under} normal glucose conditions.
However, the synchronous coordination of insulin and glucagon pulses at low glucose was \rev{more} pronounced in human islets that \rev{had} more $\alpha$ cells.
When $\beta$ cells were selectively removed to mimic diabetic conditions, the anti-synchronicity of insulin and glucagon pulses was deteriorated at high glucose, but it could be partially recovered when the re-aggregation of \rev{remaining} cells was considered.
Finally, the third cell type, $\delta$ cells, \rev{which} introduced additional complexity in the multicellular system, 
\rev{prevented the} excessive synchronization of hormone pulses.
Our computational study suggests \rev{that} controllable synchronization \rev{is} a design principle of pancreatic islets.

\section*{Author Summary}
Understanding \rev{the} design principles of living systems \rev{and their functional implications} is one of the most fundamental issues in biology.
The islet\rev{s} of Langerhans can be a good system \rev{with which} to explore design principles of multicellular systems \rev{because their} cellular components, interactions, and organizations have been largely quantified \jj{with} diabetes research.
In this study, we integrate \rev{this} information using a mathematical model 
and predict \rev{the} potential benefits of \rev{different} designs of pancreatic islets. 
We found that pancreatic islets have a special design for controlling \rev{the} synchronization of hormone pulses:
islet cells \rev{are} highly synchronized to concentrate hormone pulses \rev{under} low-/high-glucose conditions,
while they \rev{are} desynchronized to suppress unnecessary hormone actions \rev{under} normal glucose conditions.
The flexibility of responses to external stimuli is an essential feature of living systems,
and the controllability of hormone responses can provide a new \rev{perspective} on the malfunction of diabetic islets.


\section*{Introduction}
Living systems have structural designs for their functional demands, which has been referred \rev{to} as {\it symmorphosis}~\cite{Weibel91}.
The islets of Langerhans in the pancreas also have unique \rev{architecture, which helps them to accomplish their functional goal maintaining constant blood glucose.}
The multicellular system is composed mainly of three cell types: \jj{insulin-secreting $\beta$ cells, glucagon-secreting $\alpha$ cells,} 
and somatostatin-secreting $\delta$ cells.
\jj{Insulin and glucagon} are reciprocal hormones that \rev{decrease and increase} \rev{blood glucose, respectively}.
Interestingly, different species have different islet architectures~\cite{Brissova05, Cabrera06, Bosco10, Steiner10}.
Mouse islets have a shell-core structure in which $\beta$ cells are located in the core, while non-$\beta$ cells are \rev{located in the periphery, surrounding the core}.
However, large human islets, \rev{which contain} a lower fraction of $\beta$ cells, have a mixing structure in which non-$\beta$ cells are not only distributed \rev{throughout} the islet periphery but also scattered within \rev{the} islets.
Recently\rev{,} we have found that the spatial organization of islet cells \rev{follows} a conserved rule \rev{in which} homotypic cell-cell contacts have a slightly stronger attraction than heterotypic contacts~\cite{Hoang14}.

\jj{Insulin, glucagon,} and somatostatin secretions are pulsatile\rev{,} like other endocrine hormones.
The three pulses are not independent, but coordinated: \rev{t}he \jj{approximate} out-of-phase coordination of \jj{insulin and glucagon} pulses has been observed in the blood of normal humans but not in the blood of diabetic patients~\cite{Menge11, Rohrer12}.
Perifused islets have also shown out-of-phase coordination \rev{as well as} the in-phase coordination
of insulin and somatostatin at \rev{high glucose}~\cite{Hellman09, Hellman12}.
\rev{P}hase coordination implies \rev{that there is communication} between islet cells.
Indeed, \rev{this communication has} been extensively studied \rev{in the form of} paracrine/autocrine interactions \rev{via} hormones and neurotransmitters~\cite{Koh12}. 

\begin{figure}[h]
\centerline{\includegraphics[width=0.9\textwidth]{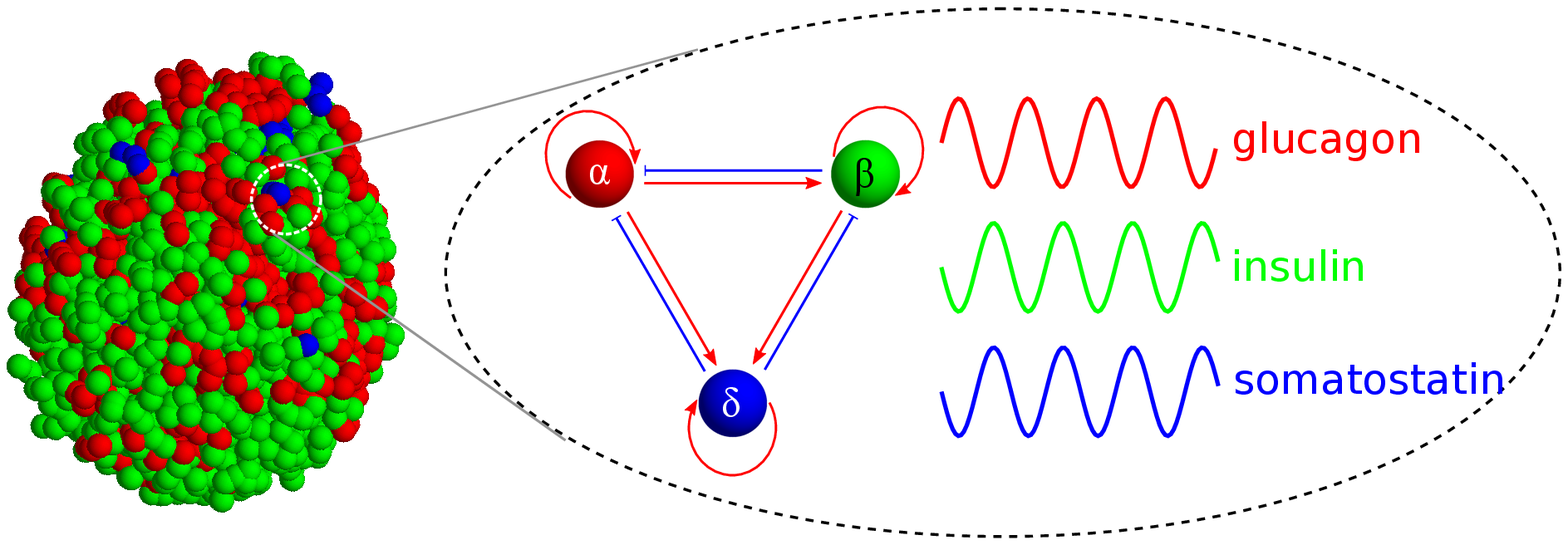}}
\caption{{\bf Cellular organization and interaction in pancreatic islets.} 
Endocrine $\alpha$ (red), $\beta$ (green), and $\delta$ \rev{cells (blue)} generate pulses of glucagon, insulin, and somatostatin, respectively.
They positively (red arrows) or negatively \rev{affect (blue bar-headed arrows)} hormone pulses of neighboring cells.}
\label{fig1}
\end{figure}

Considering the \rev{mechanism of cellular communication}, the spatial organization of islet cells should have functional implications.
\rev{Clustered $\beta$ cells secrete insulin more robustly} than single $\beta$ cells~\cite{Sherman91, Jonkers99}. 
Recent studies have demonstrated the effectiveness of $\beta$-cell clustering by systematically controlling the size of $\beta$-cell aggregates~\cite{Hraha14a, Hraha14b}.
However, to understand the  organization of $\alpha$, $\beta$, and $\delta$ cells beyond $\beta$-cell clustering, we await technical \rev{innovations} that \rev{allow the identification of} different cell types within islets and \rev{the recording of} their activities \rev{at a high} resolution.
Nevertheless, we know (i) how single $\alpha$, $\beta$, and $\delta$ cells produce hormone pulses~\cite{Hong13}\rev{,}
(ii) how the hormone pulses are affected by other hormone pulses~\cite{Hong13}\rev{,} and (iii) how those cells are spatially distributed within islets~\cite{Hoang14}.
\rev{We were motivated to integrate} the model and \rev{the} data (Fig.~\ref{fig1})\rev{,} and \rev{we} computationally \rev{inferred the} functional implications of the islet architecture.
In this computational study, we \rev{found} that the organization of islet cells and their interactions are designed to
produce synchronous hormone pulses \rev{under low- and high-}glucose conditions and asynchronous hormone pulses \rev{under} normal glucose conditions.
We also \rev{observed} that the controllable synchronization \rev{was} deteriorated in diabetic islets.

\section*{Islet model}\label{sec:model}
We formulate\rev{d} an islet model \rev{on the basis of} two observations:
\begin{itemize}
\item[(i)]
{Islet cells are intrinsic oscillators that produce pulses of endocrine hormones.}
\item[(ii)]
{Islet cells interact with neighboring cells via paracrine/autocrine signaling.}
\end{itemize}
First, insulin pulses have been widely observed \rev{in} blood samples \rev{obtained from} live animals~\cite{Lang79} \rev{as well as} perfused pancreata~\cite{Stagner80} and islets~\cite{Hellman09}.
\rev{P}ulsatility is an intrinsic property of $\beta$ cells because isolated $\beta$ cells can still generate oscillations of intracellular calcium concentration~\cite{Grapengiesser91, Tengholm09}, a trigger of insulin secretion.
Isolated $\alpha$ cells~\cite{Berts96a} and $\delta$ cells~\cite{Berts96b} can also generate calcium oscillations.
Second, islet cells secrete hormones and/or neurotransmitters, and the signaling molecules affect the hormone secretions of neighboring cells.
The paracrine/autocrine interactions between $\alpha$, $\beta$, and $\delta$ cells are summarized in Table I.

To obtain a robust conclusion independent \rev{of the details of the model}, we \rev{generated} a minimal model that incorporates the two basic observations.
The dynamics of coupled oscillators \rev{have} been extensively studied \rev{using} the prototypic model, \rev{the} Kuramoto model~\cite{Acebron05}. \rev{Thus, we adopted an} oscillator model:
\begin{equation}
       \dot{\theta}_i = \omega_{i} + \sum_{j \in \Lambda_i} K_{\sigma_i \sigma_j} \sin(\theta_j - \theta_i),
        \label{eq:model}
        \end{equation}
where $\theta_i \in \mathbb{R}$ and $\sigma_i \in \{\alpha, \beta, \delta\}$ are the phase and type of the $i$th cell among $N$ cells within an islet.
The phase represents \rev{the degree} of hormone secretion: \jj{given amplitude,} $\theta=0$ and $\theta=\pi$ represent minimal and maximal secretion, respectively.
Each cell produces oscillatory hormone secretion with \rev{an} intrinsic frequency $\omega_{i}$.
In this study, we focus\rev{ed} on the oscillation with a period of $\omega^{-1} \sim 10$ minutes.
For simplicity, we assume\rev{d} that every cell \rev{had} an identical frequency $\omega_i=\omega$; this assumption \rev{was} relaxed later. 

The second term in Eq.~(\ref{eq:model}) represents the interactions \rev{of} the nearest neighboring $j$th cells.
The neighborhood set $\Lambda_i$ of the $i$th cell \rev{was} predetermined from \rev{the} data of \rev{the} islet structures.
The \rev{strength of the interaction} from the $j$th cell to the $i$th cell is
\begin{equation}
        K_{\sigma_i \sigma_j}=  A_{\sigma_i \sigma_j} {r_{\sigma_j}}{r_{\sigma_i}}^{-1}
        \label{eq:coupling}
        \end{equation}
where $A_{\sigma_i \sigma_j}$ defines the sign of the interaction (Table 1)
and $r_{\sigma_i}$ and $r_{\sigma_j}$ represent \rev{the} relative activities of the receiver and affecter cells.
Positive/negative interactions ($A_{\sigma_i \sigma_j}=\pm1$) lead the $i$th cell to have in-phase/out-of-phase oscillations with the $j$th cell.

\begin{table}[!t]
\caption{\bf Interaction signs between islet cells. }
\begin{tabular}{|c|c|c|c|}
\hline
Symbol & Interaction & Sign & Reference \\ \hline
$A_{\alpha \alpha}$ &  $\alpha \to \alpha$ & $+$ & \cite{Cabrera08, Cho10, Leibiger12, Gilon15} \\ 
$A_{\beta \alpha}$ & $\alpha \to \beta$ & $+$ & \cite{Samols65, Kawai95, Brereton07, Rodriguez-Diaz11} \\
$A_{\delta \alpha}$ & $\alpha \to \delta$ & $+$ & \cite{Patton76, Weir78, Dolais-Kitabgi81} \\
$A_{\alpha \beta}$ & $\beta \to \alpha$ & $-$ & \cite{Cherrington76, Samols76, Rorsman89, Ishihara03, Ravier05, Franklin05, Tuduri08, Ramracheya10} \\
$A_{\beta \beta}$ &$\beta \to \beta$ & $+$ & \cite{Aspinwall99, Leibiger02, Braun10, Jacques-Silva10} \\
$A_{\delta \beta}$ & $\beta \to \delta$ & $+$ & \cite{Honey79, Bertrand90, Meulen15} \\
$A_{\alpha \delta}$ & $\delta \to \alpha$ & $-$ & \cite{Koerker74, Orci75, Cherrington76, Guillemin76} \\
$A_{\beta \delta}$ & $\delta \to \beta$ & $-$ & \cite{Koerker74, Orci75, Cherrington76, Guillemin76, Daunt06} \\
$A_{\delta \delta}$ & $\delta \to \delta$ & $\cdot$ &  $\cdot$ \\ \hline
 \end{tabular}
\label{table1}
\end{table}

The interactions are mediated \rev{by} signaling molecules secreted from $\alpha$, $\beta$, and $\delta$ cells.
Thus\rev{,} the interaction strengths should \rev{be dependent on the activities of cells that} are governed by glucose \rev{level}.
\jj{One can consider those activities as average hormone secretions at different glucose concentrations~\cite{Vieira07, Walker11}.}
In general, $\alpha$ cells are active at low glucose, while $\beta$ and $\delta$ cells are active at high glucose.
\jj{Here we included the glucose conditions implicitly in the activities of islet cells,
and simply defined low ($r_\alpha > r_\beta$), normal ($r_\alpha = r_\beta$), and high glucose conditions ($r_\alpha < r_\beta$).
Thus $r_\beta/r_\alpha$ is a proxy parameter for glucose conditions of which scale can be different from real glucose concentrations.}
The current $K_{\sigma_i \sigma_j}$ considers \rev{the} activities of \rev{both} affecter and receiver cells;
\rev{t}he pair of \rev{a} strong affecter and \rev{a} weak receiver exhibits maximal coupling.
We considered an alternative activity-dependent interaction, $K_{\sigma_i \sigma_j}= A_{\sigma_i \sigma_j} r_{\sigma_j}$,
which ignores the activity of receiver cells. \rev{Because these} two cases did not show significant differences, we focus\rev{ed} on the former definition in Eq.~(\ref{eq:coupling}).
This setting helps to reduce the number of parameters from nine $K_{\sigma \sigma'}$ to three $r_{\sigma}$.
\rev{Next,} we have a constraint, $|K_{\sigma_i \sigma_j}|=|K_{\sigma_j \sigma_i}|^{-1}$,
and this implies that every autocrine interaction has a unity of strength, $|K_{\alpha \alpha}|=|K_{\beta \beta}|=|K_{\delta \delta}|=1$.

\section*{Results}

\subsection*{Islet cells generate \rev{the} glucose-dependent coordination of insulin and glucagon pulses}
We start\rev{ed by considering} simple islets that \rev{had} only $\alpha$ and $\beta$ cells, two major \rev{cell} populations ($> 90\%$).
Considering the antagonistic roles of the two cell types for glucose homeostasis, they are likely to inhibit each other.
Unexpectedly, however, they \rev{showed} an asymmetric interaction rather than mutual inhibition:
$\beta$ cells indeed suppress\rev{ed} $\alpha$ cells ($K_{\alpha \beta} < 0$), 
but $\alpha$ cells stimulate\rev{d} $\beta$ cells ($K_{\beta \alpha} > 0$).
We examined how the asymmetric interaction affected the coordination of hormone pulses within islets.

\begin{figure}
\centerline{\includegraphics[width=0.9\textwidth]{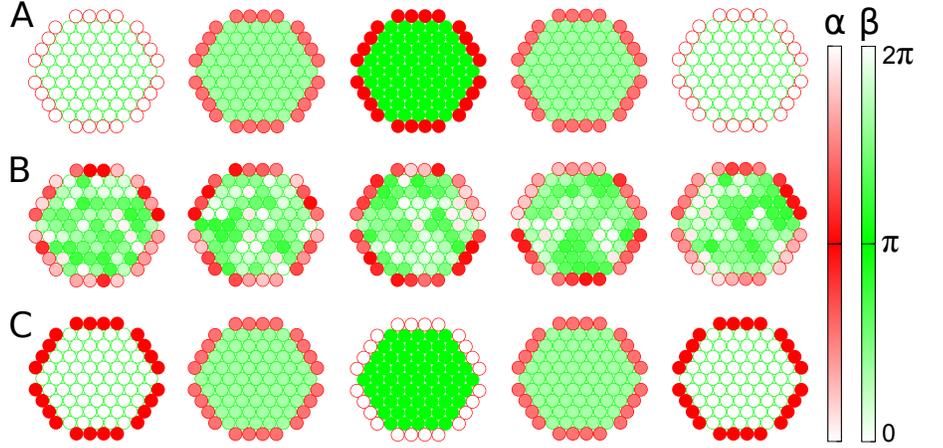}}
\caption{{\bf Snapshots of islet-cell activities.}
Sequential phase changes of $\alpha$ (red circle) and $\beta$ \rev{cells (green circle)} with time at different glucose conditions:
(A) $r_\beta/r_\alpha=0.1$ (low glucose); (B) $r_\beta/r_\alpha=1$ (normal glucose); (C) $r_\beta/r_\alpha=10$ (high glucose).
Each cell spontaneously alternates its phase between $0$ (light color) and $\pi$ (dark color),
and its neighboring cells \rev{perturb} the oscillation. Note that cross sections of three-dimensional structures are displayed for clarity.
}
\label{fig2}
\end{figure}

First, we simulated the dynamics of $\alpha$ and $\beta$ cells in Eq.~(\ref{eq:model}) with a prototypic organization of core $\beta$ cells and peripheral $\alpha$ cells.
The multicellular system had three equilibrium states\rev{,} depending on \rev{the} glucose conditions:
\begin{itemize}
\item[(i)] {In-phase synchronous state.} When $\alpha$ cells \rev{were} active ($r_{\alpha} > r_{\beta}$) at low glucose, they secrete\rev{d} 
\jj{neurotransmitters, which sensitize\rev{d} $\beta$ cells to secrete insulin~\cite{Rodriguez-Diaz11}.}
Here the positive interaction ($K_{\beta \alpha}=r_{\alpha}/r_{\beta}$) from $\alpha$ to $\beta$ cells dominate\rev{d} the negative interaction ($K_{\alpha \beta}=-r_{\beta}/r_{\alpha}$) from $\beta$ to $\alpha$ cells. 
In addition, the positive autocrine interactions ($K_{\alpha \alpha}=K_{\beta \beta}=1$) help\rev{ed} homotypic cells \rev{synchronize with} each other.
Given these conditions, synchronous $\alpha$ cells \rev{were} coordinated in phase with synchronous $\beta$ cells (Fig.~\ref{fig2}A and Video S1A).
This state represent\rev{ed} the in-phase coordination of glucagon and insulin pulses. 
\item[(ii)] {Asynchronous state.} When $\alpha$ and $\beta$ cells \rev{were} equally active ($r_{\alpha} = r_{\beta}$) at normal glucose, 
the asymmetric interaction \rev{had} the same strength ($|K_{\beta \alpha}| = |K_{\alpha \beta}|$=1). 
\rev{The} $\alpha$ cells \rev{were ``confused" about whether to be active} 
because neighboring $\alpha$ cells activate\rev{d} them but neighboring $\beta$ cells equally suppress\rev{ed} them.
This \rev{incongruous} condition ultimately \rev{resulted in} both $\alpha$ and $\beta$ cells \rev{becoming} asynchronous, although local homotypic clusters temporally showed synchronous behaviors (Fig.~\ref{fig2}B and Video S1B).
\rev{L}ocal synchronization has been observed in a recent experimental study~\cite{Almaca15}.
\item[(iii)] {Out-of-phase synchronous state.} When $\beta$ cells \rev{were} active ($r_{\alpha} < r_{\beta}$) at high glucose, 
they secrete\rev{d} insulin \jj{and neurotransmitters}, which suppress\rev{ed} $\alpha$ cells \rev{from secreting} glucagon.
Unlike the low-glucose condition, 
the negative interaction ($K_{\alpha \beta}=-r_{\beta}/r_{\alpha}$) from $\beta$ to $\alpha$ cells dominate\rev{d} the positive interaction 
($K_{\beta \alpha}=r_{\alpha}/r_{\beta}$) from $\alpha$ to $\beta$ cells. 
\rev{Thus,} the synchronous $\alpha$ cells \rev{were} coordinated out of phase with the synchronous $\beta$ cells (Fig.~\ref{fig2}C and Video S1C).
This state represent\rev{ed} the out-of-phase coordination of glucagon and insulin pulses\rev{,} which has been repeatedly reported~\cite{Menge11, Rohrer12, Hellman09, Hellman12}.
\end{itemize}

The three states were not sharply divided but smoothly \rev{altered} with glucose conditions ($r_{\beta}/r_{\alpha}$).
\rev{Thus}, we introduced order parameters $R_{\alpha}$ and $R_{\beta}$ that measured the \rev{degree of synchronization 
between $\alpha$ cells and $\beta$ cells}, respectively (See Materials and Methods):
$R_{\alpha}=1$ and $0$ represent perfect synchronization and desynchronization between $\alpha$ cells, \rev{respectively.} The same is true for $R_{\beta}$.
In addition, we measured the phase difference $\Delta \Theta$ between average $\alpha$-cell phase $\Theta_\alpha$ and $\beta$-cell phase $\Theta_\beta$: $\Delta \Theta=0$ and $\pi$ represent perfect in-phase and out-of-phase states, \rev{respectively}.

\begin{figure}
\centerline{\includegraphics[width=1.0\textwidth]{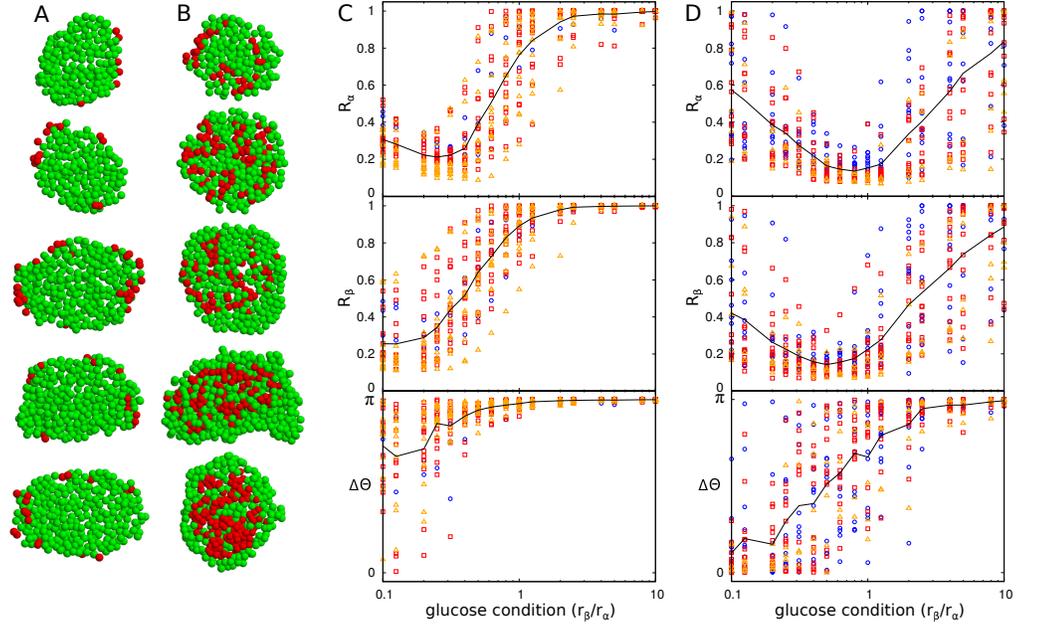}}
\caption{{\bf Glucose-dependent synchronization of islet cells in mouse and human islets.}
Cross sections of (A) mouse and (B) human islets with $\alpha$ (red) and $\beta$ \rev{cells (green)}.
Synchronization and phase coordination of islet cells in (C) mouse \rev{(n=29) and (D) human islets (n=28)}.
Synchronization indices $R_{\alpha}$ and $R_{\beta}$ represent the degrees of synchronization between $\alpha$ cells
and between $\beta$ cells, respectively, and phase index $\Delta \Theta$ indicates the difference of average phases
of $\alpha$ and $\beta$ cells.
Islets are categorized into three groups according to size $N$: small islets ($N < 1000$ cells, blue circle);
medium islets ($1000 < N < 2000$, red square); and large islets ($N > 2000$, orange triangle). 
Black lines represent average values of corresponding indices of every islet.
}
\label{fig3}
\end{figure}

Using these order parameters, we examined the dynamics of $\alpha$ and $\beta$ cells 
that interact\rev{ed} given \rev{the} spatial distributions in mouse and human islets (Figs.~\ref{fig3}A and \ref{fig3}B).
We observed the above three states \rev{frequently} in the two species.
However, the out-of-phase synchronous state was more pronounced in mouse islets that \rev{had} more $\beta$ cells (Fig.~\ref{fig3}C),
while the in-phase synchronous state was \rev{more} pronounced in human islets that \rev{had} more $\alpha$ cells (Fig.~\ref{fig3}D).
Different sizes of islets showed similar dynamics in both mouse and human islets.
The size independence is of particular interest in human islets that have different cellular compositions \rev{according to} their size.
\rev{Interestingly}, $\alpha$ cells and $\beta$ cells were partially synchronous ($R_{\alpha}<1$ and $R_{\beta} < 1$)\rev{,} 
except \rev{under} very high\rev{-}glucose conditions.
\jj{In particular, the in-phase synchronous state at low glucose was largely suppressed in mouse islets.
\rev{A}synchronous oscillation of $\alpha$ cells at low glucose has been experimentally observed~\cite{Nadal99}.
} 

Pancreatic islets \rev{contain other} cell types. 
Although \rev{$\delta$ cells compose a minor portion of the} population ($< 10 \%$), they can affect the cellular dynamics
because they interact with $\alpha$ and $\beta$ cells (Table 1), \rev{as}
$\delta$ cells suppress both $\alpha$ and $\beta$ cells, \rev{but} are activated by both $\alpha$ and $\beta$ cells.
To probe the role of $\delta$ cells, we compared the cellular dynamics in the presence and absence of $\delta$ cells within islets (Fig. S1).
The complex interactions between $\alpha$, $\beta$, and $\delta$ cells \rev{disrupted} the synchronizations of $\alpha$ and $\beta$ cells.
In general, the existence of $\delta$ cells decreased the degree of synchronization, but the minor population did not
dramatically modify the above results that ignored $\delta$ cells.
Thus, \rev{for simplicity, we do not consider $\delta$ cells hereafter}.

\subsection*{The \jj{organizations and interactions of islet cells are} designed for smooth transitions between synchronous and asynchronous hormone pulses}
\begin{figure}
\centerline{\includegraphics[width=0.6\textwidth]{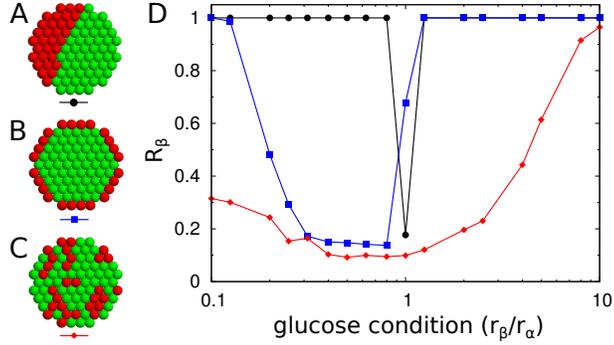}}
\caption{{\bf Islet structure and synchronization.}
(A) Complete sorting (black circle), (B) shell-core sorting (blue square), and (C) mixing (red diamond) structures of $\alpha$ (red) and $\beta$ \rev{cells (green)}. 
\rev{The} total number of cells and the fraction of $\beta$ cells are fixed as $N=725$ and $p_\beta = 0.6$\rev{, respectively} for \rev{all} three structures.
Note that cross\rev{-}sections of three-dimensional structures are displayed for clarity.
(D) Synchronization index $R_\beta$ of $\beta$ cells for different glucose conditions.
}
\label{fig4}
\end{figure}

To systematically investigate the design principles of natural islets,
we considered artificial islets that \rev{had} different organizations of islet cells or different interactions between them.
As a backbone for \rev{the} three-dimensional arrangement of islet cells, we adopted hexagonal-close-packed lattices~\cite{Hoang14, Nittala07}
and controlled the spatial distributions and compositions of $\alpha$ and $\beta$ cells, \rev{($p_\alpha$ and $p_\beta$, respectively).}
We \rev{generated} different islet structures by tuning the relative adhesions between cell types (See Materials and Methods).
Three distinct structures \rev{were} 
(i) \rev{the} complete sorting structure, \rev{in which} homogeneous cell clusters \rev{were} divided into left and right hemispheres; 
(ii) \rev{the} shell-core sorting structure, \rev{in which} $\beta$ cells \rev{were} clustered in the core and $\alpha$ cells \rev{were} in the periphery;
and (iii) \rev{the} mixing structure, \rev{in which} $\alpha$ and $\beta$ cells \rev{were} intermingled \rev{with} each other. 
For a fixed cellular composition \rev{($p_\alpha=0.4, p_\beta=0.6$)}, the three structures showed different patterns of synchronization (Fig.~\ref{fig4}).
Unlike the shell-core and mixing structures, the complete sorting structure always generated perfect synchronization between cells
except \rev{under} a very narrow glucose condition ($r_{\beta}/r_{\alpha}=1$). 
The lack of partial synchronization \rev{resulted in} abrupt transitions between the synchronous and asynchronous states. 

\begin{figure}
\centerline{\includegraphics[width=1.0\textwidth]{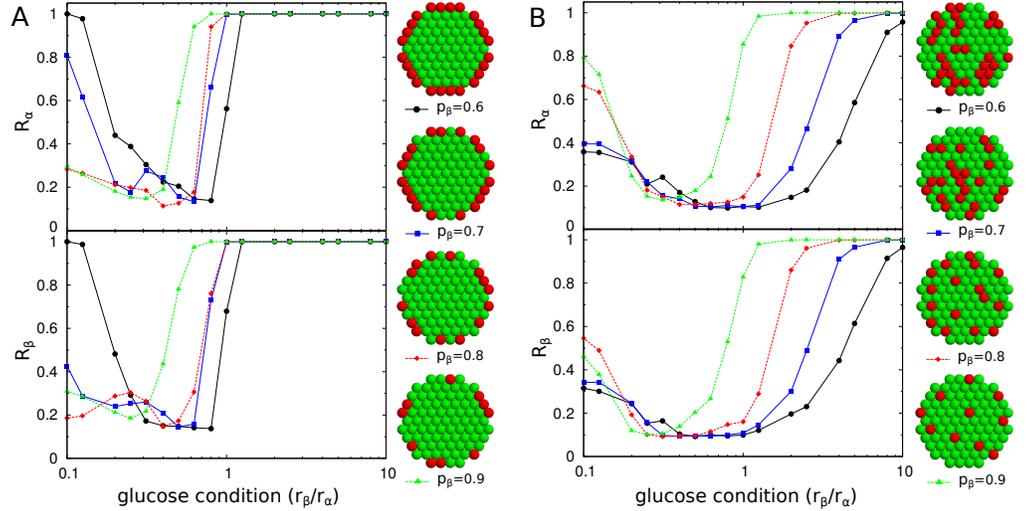}}
\caption{{\bf Cellular composition and synchronization.}
Synchronization indices $R_{\alpha}$ of $\alpha$ cells (red) and $R_{\beta}$ of $\beta$ cells (green)
for various cellular compositions in (A) shell-core sorting and (B) mixing structures.
The fractions of $\beta$ cells are $p_{\beta}=0.6$ (black circle), $0.7$ (blue square), $0.8$ (red diamond), and $0.9$ (green triangle)
among $N=725$ cells.
Note that cross\rev{-}sections of three-dimensional structures are displayed for clarity.
}
\label{fig5}
\end{figure}

\rev{Next,} we controlled cellular compositions given a total cell number ($N=725$).
Mouse islets have \rev{a} shell-core structure with a high fraction of $\beta$ cells \rev{($p_{\beta} \approx 0.9$)}.
If the $\beta$-cell fraction was decreased in mouse islets, \rev{then the multi}-cellular dynamics showed \rev{an} enhancement of the in-phase synchronous state and sharper slopes of $R_{\alpha}$ and $R_{\beta}$ \jj{with respect to the glucose conditions} (Fig.~\ref{fig5}A).
However, human islets have the mixing structure with a smaller fraction of $\beta$ cells \rev{($p_{\beta} \approx 0.6$).}
If the $\beta$-cell fraction was increased in human islets like the fraction in mouse islets, \rev{then}
the modified human islets had \rev{the} enhanced in-phase synchronization and sharper slopes of $R_{\alpha}$ and $R_{\beta}$ (Fig.~\ref{fig5}B).
Therefore, the large $\beta$-cell fraction in mouse islets and small $\beta$-cell fraction in human islets
are effective \rev{in suppressing} the in-phase synchronization between insulin and glucagon pulses,
and \rev{in preventing} sharp transitions between the synchronous and asynchronous hormone pulses as glucose conditions change.
These conclusions were the same for different sizes of islets (Fig. S2).

The opposite dependence of $\beta$-cell fractions for mouse and human islets was related 
to the number and size inhomogeneity of $\alpha$-cell clusters.
\rev{T}he number of $\alpha$ cells was not sufficient to form a large cluster in the shell-core structure, \rev{so}
several clusters of $\alpha$ cells \rev{that varied in size} existed. 
\rev{However}, as the number of $\alpha$ cells increased in the mixing structure, 
clusters of $\alpha$ cells \rev{of various sizes started to appear}.
The separate clusters of $\alpha$ cells contributed to \rev{the diminishing of} the synchronization of $\alpha$ cells.

\begin{figure}
\centerline{\includegraphics[width=0.9\textwidth]{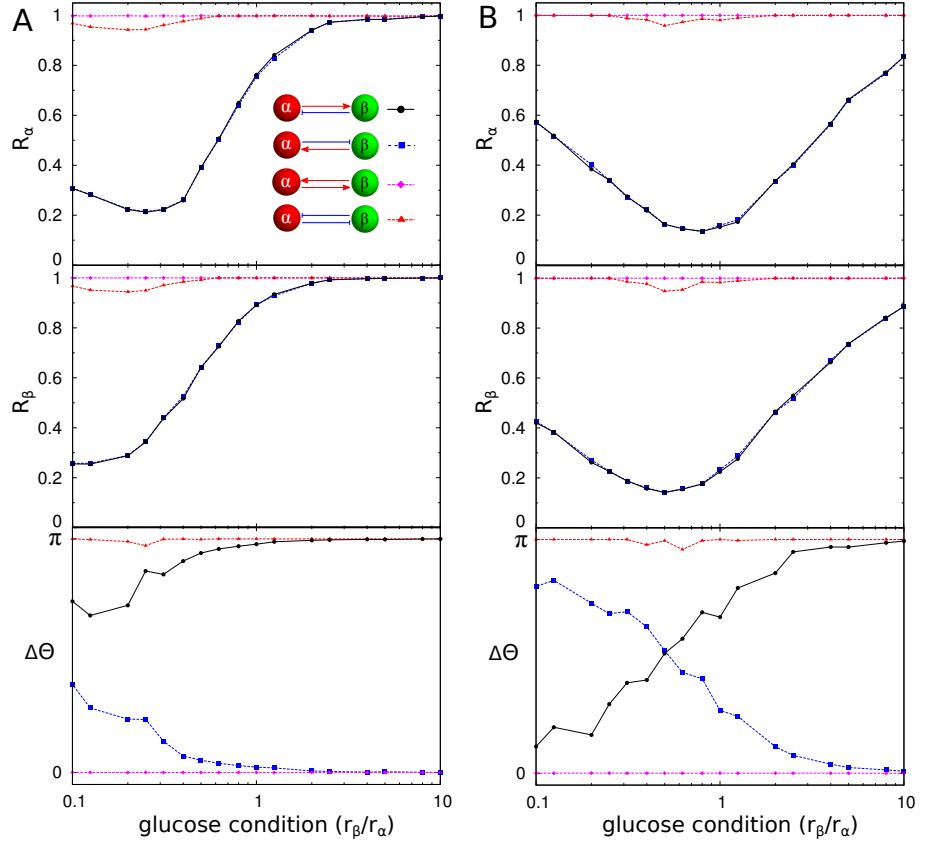}}
\caption{{\bf Cellular interaction and synchronization.}
Synchronization indices $R_{\alpha}$ of $\alpha$ cells (red) and $R_{\beta}$ of $\beta$ cells (green)
and the average phase difference  $\Delta \Theta$ between $\alpha$ and $\beta$ cells
are measured for four scenarios of the mutual interaction between $\alpha$ and $\beta$ cells in (A) mouse and (B) human islets:
(i) $\alpha$ cells activate $\beta$ cells, while $\beta$ cells suppress $\alpha$ cells (black circle);
(ii) opposite interaction to (i) (blue square);
(iii) mutual activation (magenta diamond);
and (iv) mutual inhibition (red triangle).
Note that (i) black line represents the result from the true interaction in natural islets (See Fig.~\ref{fig3}).
}
\label{fig6}
\end{figure}

Next\rev{,} we modified the interactions between $\alpha$ and $\beta$ cells by considering every possibility \rev{in regard to their mutual interaction}.
In natural islets, $\alpha$ cells \jj{sensitize} $\beta$ cells, while $\beta$ cells suppress $\alpha$ cells.
Unlike the asymmetric interaction, when the two cells symmetrically activated or inhibited each other,
they always generated synchronous hormone pulses \rev{that were} independent of \rev{the} glucose conditions (Fig.~\ref{fig6}).
However, the mutual activation model always showed in-phase coordination of insulin and glucagon pulses, 
while the mutual inhibition model always showed out-of-phase coordination of \rev{the pulses}.
When the asymmetric interaction was reversed, the controllability of \rev{the} synchronization was intact,
but the phase coordination of insulin and glucagon pulses was reversed for \rev{the} glucose conditions.
If $\alpha$ cells \jj{inhibited} $\beta$ cells \rev{and} $\beta$ cells activated $\alpha$ cells, \rev{then}
they could generate \rev{the} in-phase coordination of insulin and glucagon pulses \rev{under} high\rev{-}glucose conditions.

\subsection*{Diabetic islets fail to produce coordinated pulses of insulin and glucagon}
\begin{figure}
\centerline{\includegraphics[width=0.8\textwidth]{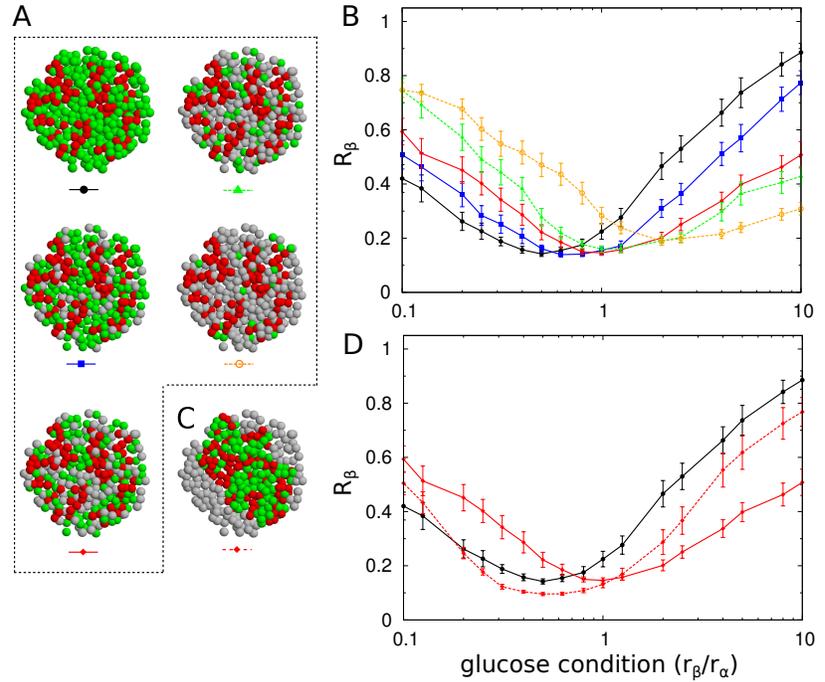}}
\caption{{\bf Synchronization of islet-cells under $\beta$-cell loss.}
(A) To simulate human diabetic islets, $\beta$ cells \rev{were} selectively removed randomly from human islets (n=28): no (black circle), $30\%$ (blue square), $50\%$ (red diamond), $70\%$ (green triangle), and $90\%$ \rev{loss of $\beta$-cell mass (yellow empty circle)}.
Note that cross\rev{-}sections of three-dimensional structures are displayed for clarity.
(B) Synchronization index $R_{\beta}$ of $\beta$ cells for the loss of $\beta$ cells.
(C) The \rev{remaining} cells after the \rev{removed $\beta$ cells ($50\%$) were} re-aggregated.
(D) Synchronization index $R_{\beta}$ of $\beta$ cells with (red diamond, solid line) 
and without (red diamond, dashed line) the consideration of re-aggregation.
The error bars represent standard errors.
}
\label{fig7}
\end{figure}

We \rev{simulated} diabetic islets by removing $\beta$ cells (Fig.~\ref{fig7}A).
The random removal of $\beta$ cells attenuated the synchronization of islet cells \rev{under} high\rev{-}glucose conditions (Fig.~\ref{fig7}B).
In particular, a drastic change was found \rev{when approximately 50\% of the $\beta$ cells were moved}.
\rev{A} loss of out-of-phase coordination of insulin and glucagon pulses has been observed in diabetic patients~\cite{Menge11}.
Furthermore,  $\alpha$ cells became relatively abundant due to the selective loss of $\beta$ cells.
This change \rev{strengthened} the positive interactions from $\alpha$ cells, 
which then enhanced \rev{the} synchronization of islet cells \rev{under} low\rev{-}glucose conditions.

To \rev{more realistically} simulate the diabetic islets, \rev{the} re-aggregation of \rev{remaining} cells should be considered (Fig.~\ref{fig7}C),
because the sites of removed $\beta$ cells could not \rev{physically remain as an} empty space.
One interesting question is whether the space\rev{-}filling is just a passive process
or \rev{whether} it can be an active process \rev{used} to escape the deteriorated condition.
When we considered re-aggregation (See Materials and Methods),
the enhanced synchronization at low glucose and suppressed synchronization at high glucose
were partially recovered to original patterns for normal islets (Fig.~\ref{fig7}D).
This \rev{finding} suggested that the re-organization of islet cells under \rev{diabetic conditions could actively contribute to} recovery.

\section*{\rev{Discussion}}
We computationally studied the design principles of pancreatic islets by integrating their structure and function \rev{in a model}.
Our model incorporated \rev{the} pulsatility of hormone secretions, paracrine/autocrine interactions, 
and spatial organization of pancreatic $\alpha$, $\beta$, and $\delta$ cells.
We found that the \rev{multicellular} system functions not only to produce synchronous hormone pulses at low/high glucose 
but also to produce asynchronous hormone pulses \rev{under} normal glucose conditions.
The controllable synchronization effectively \rev{enhanced} and suppress\rev{ed} hormone actions\rev{,} depending on \rev{the} glucose conditions.
Thus\rev{,} we proposed that the defective controllability of hormone pulses could contribute \rev{to metabolic diseases, such as} diabetes.

We predicted \rev{the} glucose-dependent \rev{coordination} of hormone pulses.
Regarding the phase coordination of \jj{insulin and glucagon} pulses, previous studies \rev{have} reported
out-of-phase~\cite{Menge11, Hellman09}, in-phase~\cite{Lang82, Jaspan86}, and no~\cite{Hansen82, Stagner80} relationships between the two pulses.
\rev{C}ontroversies could originate from variations in animal species, islet preparation, and/or experiment conductors.
Our finding, however, implies that the different phase coordinations could be partially explained by \rev{the} different glucose conditions 
in the studies. 

We adopted a minimal model with essential ingredients to reduce unnecessary complexities and obtain robust conclusions.
Indeed\rev{,} we \rev{confirmed} that some possible modifications did not change our conclusion about controllable synchronization. First, we used an alternative form of paracrine/autocrine interactions ($K_{\sigma_i \sigma_j}=A_{\sigma_i \sigma_j} r_{\sigma_j}$)  
that ignore\rev{d} the activity of receiver cells (Fig. S3A).
Second, we relaxed the assumption that every cell ha\rev{d} the same intrinsic frequencies 
by introducing some variations of $w_i \in [0.8, 1.2]$ (Fig. S3B).
Third, we applied a stronger interaction ($K_{\beta \beta}=2$) between $\beta$ cells as \rev{the} simplest way to consider their gap-junctional coupling in addition to their autocrine interaction (Fig. S3C).   
Finally, we ignored the autocrine interaction of $\delta$ cells ($K_{\delta \delta}=0$), because \rev{this} interaction has not been observed yet;
its contribution is expected to be negligible because \rev{contact} between $\delta$ cells \rev{is} very rare (Fig. S3D).

Nevertheless, the minimal model \rev{was} limited to incorporat\rev{ing} all of the observed complexities \rev{of} the system.
In the phase-oscillator model, we simplified the shape of hormone pulses as a sine wave\rev{,} although their ridge and valley durations \jj{could} depend on \rev{the} glucose conditions.
Next\rev{,} we reduced the number of parameters that describe\rev{d} the strengths of \rev{the} paracrine/autocrine interactions
by assuming that they \rev{were} governed by the activities of \rev{the} cells.
The reduced degrees of freedom may \rev{have constrained} the multicellular system \rev{from generating} richer dynamics.
\jj{Thus we cannot rule out the possibility that $\delta$ cells play a crucial role in the multicellular system.}
For example, the out-of-phase coordination of glucagon and insulin pulses \rev{at high glucose}
may be predominantly \rev{led} by the inhibitory interaction from somatostatin rather than insulin~\cite{Gylfe14, Li15}.
\jj{Third}, the paracrine interactions may depend on external glucose stimuli as well as the activities of cells.
\jj{The stimulatory interaction from $\alpha$ cells to $\beta$ cells via glucagon functions in the presence of glucose stimuli~\cite{Liu96}.
Forth, the sign of the interactions may not be fixed. Indeed, the sign of the autocrine interaction of $\beta$ cells remains controversial~\cite{Braun12}.
} 
Finally, although we focused on the short-range interactions of islet cells with no time delay,
they may have long-range interactions via blood vessels and nerves densely innervated in islets.
However, current experimental data are not comprehensive enough to probe the relevance of these complexities 
\rev{to} the controllability of \rev{a} multicellular system.

In living systems, desynchronization is \rev{as} important as synchronization~\cite{Glass01, Louzada12}.
Pancreatic islets showed one possible design for controllable synchronization.
The design principle can be applied to other multicellular systems such as neural networks 
that have both excitatory and inhibitory connections.

\section*{Materials and methods}

\subsection*{Quantification of synchronization}
The degree of synchronization between cells in the same population is characterized by a generalized order parameter~\cite{Pikovsky01}:
\begin{equation}
\label{eq:R}
R_{\sigma} e^{i 2\Theta_{\sigma}} = \frac{\sum_{j=1}^{N} \delta_{\sigma, \sigma_j} e^{i 2\theta_j}}{\sum_{j=1}^N \delta_{\sigma, \sigma_j}},
\end{equation}
where the amplitude $R_\sigma~(0 \leq R_\sigma \leq 1)$ measures the phase coherence of $\sigma \in \{\alpha, \beta, \delta\}$ cells,
and the phase $\Theta_\sigma$ represents the average \jj{phase} of $\sigma$ cells.
Here the Kronecker delta function, $\mathcal{\delta}_{\sigma, \sigma_j}$, represents 
that the $j$th cell contributes with $\delta_{\sigma, \sigma_j}=1$ only when its type is $\sigma$, otherwise $\delta_{\sigma, \sigma_j}=0$.
\rev{Notably,} we used a multiplication factor \rev{of} 2 for the order parameters
because the multicellular dynamics showed that cells in the same population were sometimes divided into two groups with a phase difference $\pi$
(See Text S1).
The usual order parameter without the multiplication factor could not distinguish this ordered condition from \rev{a} completely disordered one.

\subsection*{Structure of mouse and human islets}
We used the structural information of mouse and human islets from our previous study~\cite{Hoang14}.
Briefly, isolated mouse \rev{($n=29$)} and human ($n=28$) islets were stained with glucagon and insulin antibodies,
and the three-dimensional coordinations of individual cells within single islets were identified using a confocal microscope.
Then, \rev{the} neighborhood of each cell was identified by calculating \rev{the} cell-to-cell distances.
In six samples of human islets, $\delta$ cells stained with \rev{the} somatostatin antibody were identified as well as $\alpha$ and $\beta$ cells.

\subsection*{Simulation of islet organization and reorganization }
To examine design principles of \rev{the} natural organization of islet cells, we generated \rev{their} artificial organization.
We used hexagonal close-packed (HCP) lattices as a backbone structure for the artificial islet organization~\cite{Hoang14}.
The spatial organization of islet cells was determined by minimizing the total cell-to-cell adhesion energy,
\begin{equation}
E= - \frac{1}{2} \sum_{i=1}^N \sum_{j \in \Lambda_i} J_{\sigma_i \sigma_j},
\end{equation}
where $J_{\sigma_i \sigma_j}$ represents the adhesion energy for the contact of $i$th cell and $j$th cell 
with their corresponding cell types, $\sigma_i$ and $\sigma_j \in \{\alpha, \beta, \delta\}$; and
$\Lambda_{i}$ denotes the set of nearest neighboring cells of the $i$\rev{th} cell.
The adhesion model could generate various structures\rev{,} depending on the parameter set $J_{\sigma \sigma'}$:
complete sorting structure ($J_{\alpha \alpha}=1$, $J_{\alpha \beta}=0.5$, $J_{\beta \beta}=1$),
shell-core sorting structure ($J_{\alpha \alpha}=1$, $J_{\alpha \beta}=1.5$, $J_{\beta \beta}=3$), 
and mixing structure ($J_{\alpha \alpha}=1$, $J_{\alpha \beta}=0.98$, $J_{\beta \beta}=1$), given \rev{a} total $N=725$ cells~\cite{Hoang14}.
Briefly, we (i) randomly distributed \rev{the} numbers of $\alpha$ and $\beta$ cells on HCP lattices; 
(ii) randomly chose two cells to swap, and calculated the total adhesion energies, $E$ and $E'$, before and after exchanging their positions; 
(iii) accepted the exchange with the probability, $\min[1, e^{(E-E')/T}]$, following the Metropolis algorithm~\cite{Metropolis49},
where the parameter $T=0.2$ controls the fluctuation of cellular organizations;
and (iv) repeated these procedures in several million Monte-Carlo steps per cell to obtain an equilibrium structure.

We applied the adhesion model to simulate the re-aggregation of \rev{the remaining} cells in diabetic islets.
We considered empty sites of removed $\beta$ cells as $\bar{\beta}$ cells.
Then, the adhesion parameters $J_{\alpha \alpha}=J_{\beta \beta}=1$, $J_{\alpha \beta}=0.98$, 
and $J_{\alpha {\bar{\beta}}} = J_{\beta {\bar{\beta}}}=J_{\bar{\beta} {\bar{\beta}}}=0$
could simulate the re-aggregation in diabetic islets because cells prefer contacting with cells to contacting with empty sites (See Video S2).

\subsection*{Numerical integration}
To integrate Eq.~(\ref{eq:model}), we used the \rev{Euler method}~\cite{Press92} with a sufficiently-small time step $\Delta t = 0.01$. 
The intrinsic frequency could be set to $\omega = 0$ for convenience because the multicellular dynamics is invariant under the transformation, $\theta_i(t) - \omega t \rightarrow \theta_i(t)$.
The initial phases $\theta_i(0) \in  \mathbb{R}$ were randomly chosen.
The quantities of interest\rev{,} such as the complex order parameters\rev{,} were measured after a sufficient\rev{ly} long transient \rev{was} discarded.

\section*{Supporting Information}
\subsection*{S1 Video}
{\bf Dynamics of islet-cell activities.} 
(A) Low\rev{-}glucose condition ($r_\beta/r_\alpha=0.1$),
(B) normal glucose condition ($r_\beta/r_\alpha=1$),
and (C) high\rev{-}glucose condition ($r_\beta/r_\alpha=10$).
Activities (or phases) of $\alpha$ (red circle) and $\beta$ \rev{cells} (green circle) change with time.
Each cell spontaneously alternates its phase between 0 (light color) and $\pi$ (dark color),
and its neighboring cells perturbs the oscillation given cellular interactions.
Note that cross\rev{-}sections of three-dimensional structures are displayed for clarity.

\subsection*{S2 Video}
{\bf Re-aggregation simulation of islet cells.}
Under 50\% removal of $\beta$ \rev{cells} (green), 
the \rev{remaining} $\alpha$ (red) and $\beta$ cells re-aggregate given preferential contacts between cells
(See Materials and Methods).
Note that this simulation was conducted on a two-dimensional lattice for clarity, 
but the re-aggregation simulations for Figs.~\ref{fig7}C and \ref{fig7}D were conducted on three-dimensional lattices. 

\subsection*{S1 Text} \label{S1Text}
{\bf Model of four coupled oscillators.} 
We introduce a simple case where cells in the same population can be divided into two groups with a phase difference $\pi$
under multicellular dynamics in Eq.~\ref{eq:model}.

\subsection*{S1 Fig} 
{\bf Role of $\delta$ cells for islet-cell synchronization.}
An islet structure in (A) \rev{the} presence and (B) \rev{absence} of $\delta$ cells.
Note that cross sections of three-dimensional structures are displayed for clarity.
(C) Synchronization index $R_\beta$ of $\beta$ cells is plotted in the presence (black filled circle) and \rev{absence} (blue empty circle) of $\delta$ cells.
The error bars represent standard errors of the mean (n=6).
For the simulation, $r_\alpha=r_\delta=1$ and \rev{$r_\beta \in [0.1, 10]$} \rev{were} used.

\subsection*{S2 Fig}
{\bf Islet size and synchronization.}
Synchronization index $R_\beta$ of $\beta$ cells for three islet sizes with shell-core (left column) and mixing \rev{structures (right)}:
(A) and (B) $N=725$ (top row), (C) and (D) $1357$ (middle), and (E) and (F) $2493$ (bottom) hexagonal-close-packed lattices.
Different cellular compositions are considered as Fig.~\ref{fig5}.
The fractions of $\beta$ cells are $p_\beta =0.6$ (black circle), 0.7 (blue square), 0.8 (red diamond), and 0.9 (green triangle).
The data result\rev{ed} from averages of \rev{five} ensembles using different initial conditions for solving Eq.~\ref{eq:model}.

\subsection*{S3 Fig} 
{\bf Model robustness.} 
Synchronization index $R_\beta$ of $\beta$ cells \rev{was} examined under modifications (blue empty circle, dotted line) of the original model (black filled circle, solid line).
(A) Strength of cellular interactions, $|K_{\sigma \sigma'}| = r_\sigma'$ vs. $|K_{\sigma \sigma'}| = r_{\sigma'}/r_{\sigma}$ (Fig.~\ref{fig3}D).
(B) Intrinsic frequency, $\omega_i = [0.8, 1.2]$ vs. $\omega_i = 1$ (Fig.~\ref{fig3}D). 
(C) Stronger interaction between $\beta$ cells, $K_{\beta \beta} = 2$ vs. $K_{\beta \beta}=1$ (Fig.~\ref{fig3}D). 
(D) No interaction between $\delta$ cells, $K_{\delta \delta}=0$ vs. $K_{\delta \delta}=1$ (Fig. S1C).
 
\section*{Acknowledgments}
We thank A. Tengholm, E. Gylfe, B. Hellman, A. Sherman, and V. Periwal for helpful discussions and critical reading of the manuscript.
This research was supported in part by Basic Science Research Program
through the National Foundation of Korea funded by the Ministry
of Science, ICT \& Future Planning (No. 2013R1A1A1006655) 
and by the Max Planck Society, the Korea Ministry of Education, Science and Technology,
Gyeongsangbuk-Do and Pohang City (J.J.), 
and DK-020595 to the University of Chicago Diabetes Research and Training Center (Animal Models Core), 
DK-072473, AG-042151, and a gift from the Kovler Family Foundation (M.H.).
 
\bibliography{islet}

\newpage
\appendix
\renewcommand\thefigure{\thesection A.\arabic{figure}} 
\renewcommand\theequation{\thesection A.\arabic{equation}} 
\section*{Text S1}
\subsection*{Model of four coupled oscillators}
We considered a simple setup to demonstrate the importance of network structure on multicellular dynamics.
Suppose we put two types of cells on \rev{a} one-dimensional array with four sites.
Here\rev{,} individual cells generate intrinsic oscillations and affect the oscillations of their nearest neighboring cells.
The phase dynamics of the interacting cells on the four sites is described by 
\begin{eqnarray}
\dot{\theta}_1 &=& \omega + K_{\sigma_1 \sigma_2} \sin(\theta_2 - \theta_1), \\
\dot{\theta}_2 &=& \omega + K_{\sigma_2 \sigma_1} \sin(\theta_1 - \theta_2) + K_{\sigma_2 \sigma_3} \sin(\theta_3 - \theta_2), \\
\dot{\theta}_3 &=& \omega + K_{\sigma_3 \sigma_2} \sin(\theta_2 - \theta_3) + K_{\sigma_3 \sigma_4} \sin(\theta_4 - \theta_3), \\
\dot{\theta}_4 &=& \omega + K_{\sigma_4 \sigma_3} \sin(\theta_3 - \theta_4),
\end{eqnarray}
where $w$ is their intrinsic frequency, $\sigma_i$ is the cell type on the $i$th site, 
and $K_{\sigma_i \sigma_j}$ represents the interaction from the $j$th cell to the $i$th cell which depends on their cell types.
\rev{Because} we are interested in \rev{the} relative phases between cells, we define them: 
$x \equiv \theta_1-\theta_2$, $y \equiv \theta_2-\theta_3$, and $z \equiv \theta_3-\theta_4$.
Then\rev{,} we obtain the equations of motion for the relative phases from the four phase equations:
\begin{eqnarray}
\label{eq:x}
\dot{x} &=& -(K_{\sigma_1 \sigma_2} + K_{\sigma_2 \sigma_1}) \sin x + K_{\sigma_2 \sigma_3} \sin y, \\
\label{eq:y}
\dot{y} &=& K_{\sigma_2 \sigma_1} \sin x - (K_{\sigma_2 \sigma_3}+K_{\sigma_3 \sigma_2}) \sin y + K_{\sigma_3 \sigma_4} \sin z, \\
\label{eq:z}
\dot{z} &=& K_{\sigma_3 \sigma_2} \sin y - (K_{\sigma_3 \sigma_4}+K_{\sigma_4 \sigma_3}) \sin z.
\end{eqnarray}
Here\rev{,} $x_*, y_*, z_* \in \{0, \pi\}$ are stationary solutions of Eqs.~(\ref{eq:x})-(\ref{eq:z}).
To examine their stabilities, we can linearize $\sin (x_* + \epsilon_x) \approx S_x \epsilon_x$ around the solutions $x_*$,
where $S_x=1$ for $x_*=0$, and $S_x=-1$ for $x_*=\pi$. 
The same is true for $y_*$ and $z_*$.
Defining a vector $\bm{\epsilon}=(\epsilon_x,  \epsilon_y,  \epsilon_z)$, 
we obtain\rev{ed} a linear equation $\dot{\bm{\epsilon}}=\bm{K} \bm{\epsilon}$ with
\begin{equation}
 \bm{K} \equiv
  \begin{bmatrix}
  -(K_{\sigma_1 \sigma_2}+K_{\sigma_2 \sigma_1}) {S_x} & K_{\sigma_2 \sigma_3}{S_y} & 0 \\
  K_{\sigma_2 \sigma_1}{S_x} & -(K_{\sigma_2 \sigma_3}+ K_{\sigma_3 \sigma_2}){S_y} & K_{\sigma_3 \sigma_4}{S_z} \\
  0 & K_{\sigma_3 \sigma_2}{S_y} & -(K_{\sigma_3 \sigma_4}+K_{\sigma_4 \sigma_3}){S_z} 
  \end{bmatrix}.
\end{equation}
\rev{Next,} we \rev{confirmed} the stabilities of the eight stationary solutions, $x_*, y_*, z_* \in \{0, \pi\}$, 
by examining the sign of the eigenvalues of the matrix $\bm{K}$.
We have two $\alpha$ cells and two $\beta$ cells with different arrangements (Fig.~A.\ref{fig1}).
The strength of cellular interactions is governed by the activities $r_{\sigma_j}$ and  $r_{\sigma_i}$ of affecter and receiver cells:
$K_{\sigma_i \sigma_j} \equiv A_{\sigma_i \sigma_j} r_{\sigma_j} r_{\sigma_i}^{-1}$.
The signs of the interactions are encoded in $A_{\alpha \alpha}= A_{\beta \beta}=1$, $A_{\alpha \beta} = -1$, and $A_{\beta \alpha}=1$.
Given \rev{the} activities $r_{\sigma_j}$ and  $r_{\sigma_i}$, different cell arrangements have different stable stationary states (Fig.~A.\ref{fig1}).

\setcounter{figure}{0}
\begin{figure} 
\centerline{\includegraphics[width=1.\textwidth]{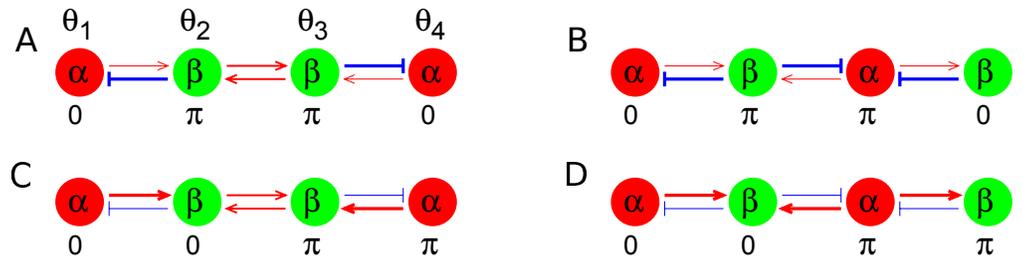}}
\caption{{\bf Synchronization of four coupled islet cells.}
(A) Shell-core and (B) mixing arrangements of two $\alpha$ (red) and two $\beta$ \rev{cells (green)}.
Numbers below cells represent stable stationary phases at \rev{high glucose} ($r_\beta/r_\alpha=10$).
Note that the phase of the first site is set to be $\theta_1=0$ for simplicity.
(C) Shell-core and (D) mixing arrangements at \rev{low glucose} ($r_\beta/r_\alpha=0.1$).
Arrows (red) represent positive interactions, while bar-headed arrows (blue) represent negative interactions.
The thickness of arrows depicts the relative strength of cellular interactions given \rev{the} glucose conditions.
}
\label{figS1}
\end{figure}
\clearpage

\newpage
\section*{Fig. S1}
\begin{figure} 
\centerline{\includegraphics[width=0.9\textwidth]{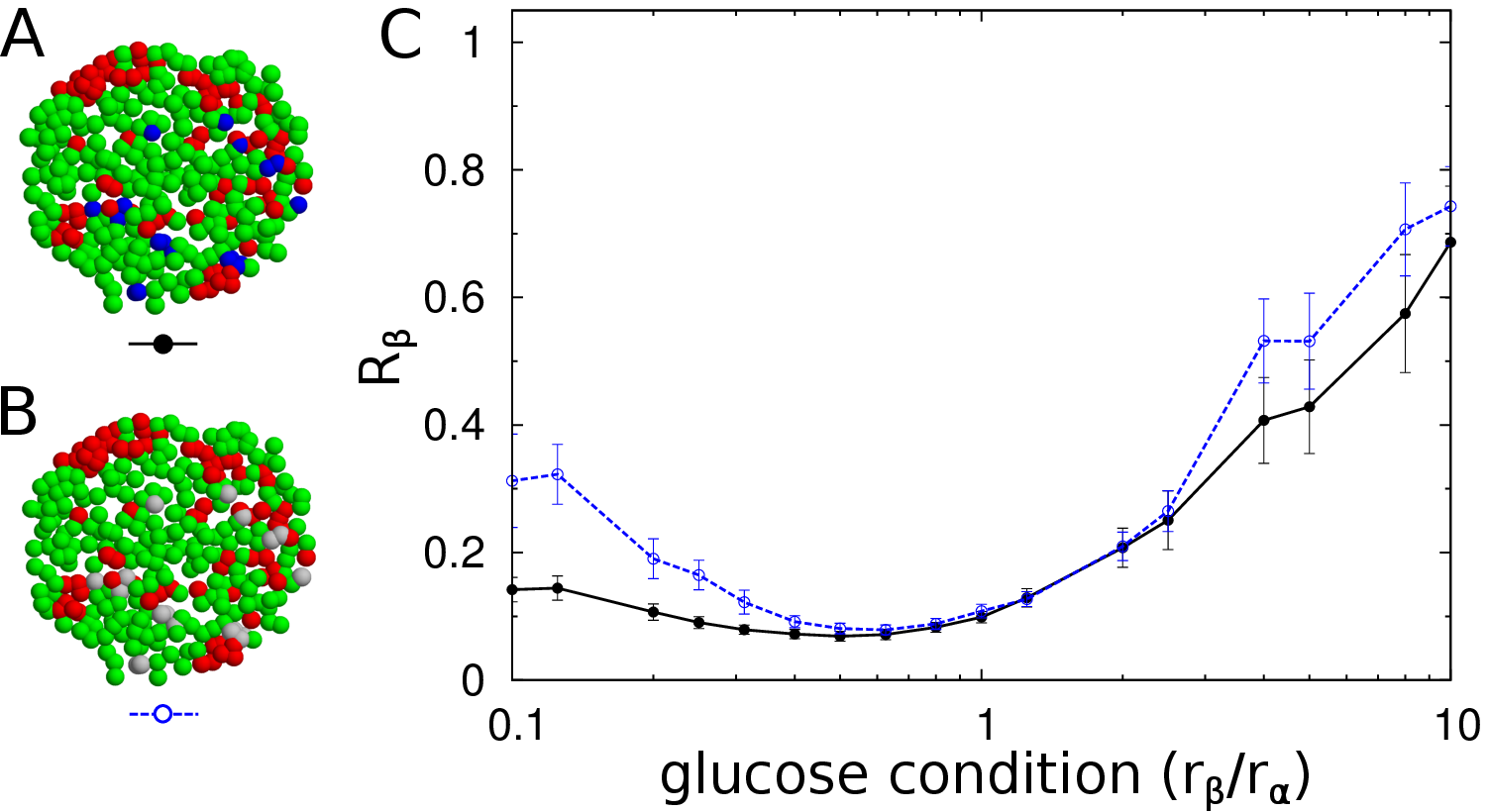}}
\end{figure}

\newpage
\section*{Fig. S2}
\begin{figure} 
\centerline{\includegraphics[width=0.9\textwidth]{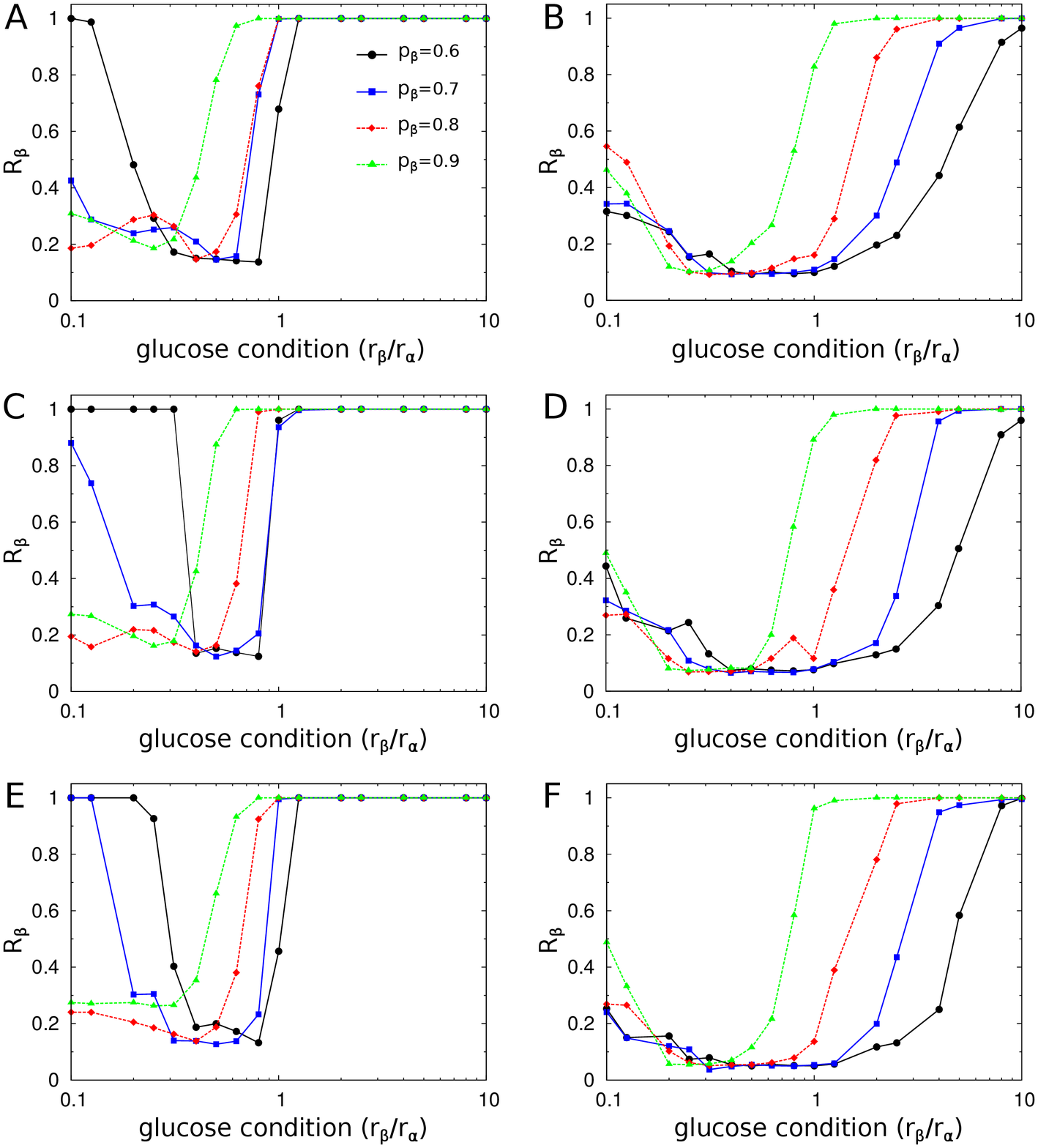}}
\end{figure}

\newpage
\section*{Fig. S3}
\begin{figure} 
\centerline{\includegraphics[width=0.9\textwidth]{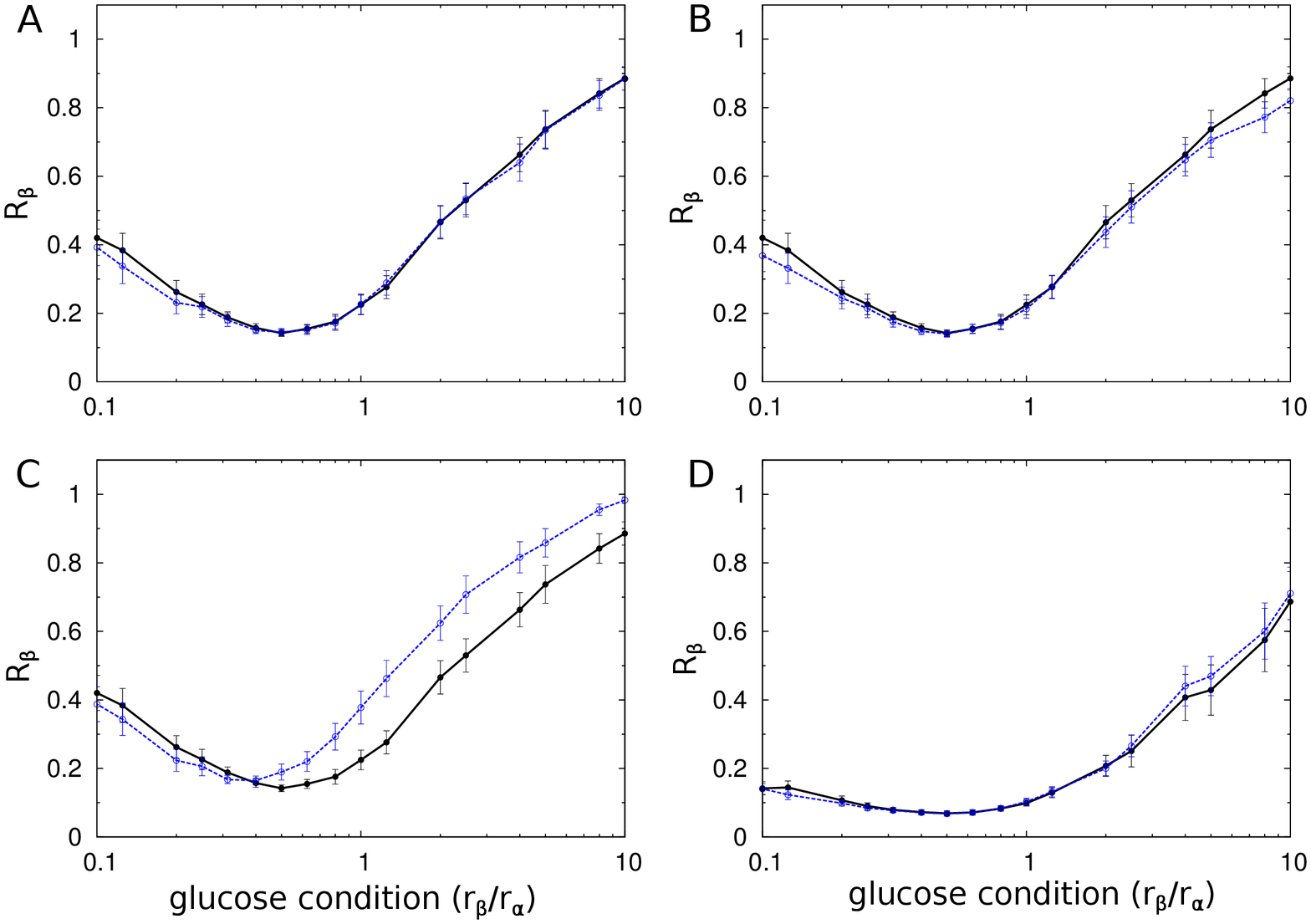}}
\end{figure}

\end{document}